\documentclass[a4paper,fleqn]{cas-sc}

\usepackage[authoryear,longnamesfirst]{natbib}
\usepackage{graphicx}
\usepackage{booktabs}
\usepackage{multirow}
\usepackage{longtable}
\def\tsc#1{\csdef{#1}{\textsc{\lowercase{#1}}\xspace}}
\tsc{WGM}
\tsc{QE}

\begin{document}
\let\WriteBookmarks\relax
\def\floatpagepagefraction{1}
\def\textpagefraction{.001}

\shorttitle{Generative AI Policies in Higher Education}    

\shortauthors{Jin et al.}  

\title [mode = title]{Generative AI in Higher Education: A Global Perspective of Institutional Adoption Policies and Guidelines}  



%

\author[1]{Yueqiao Jin}[orcid=0009-0003-7309-4984]

\cormark[1]

\fnmark[1]

\ead{ariel.jin@monash.edu}



\affiliation[1]{organization={Centre for Learning Analytics at Monash, Faculty of Information Technology, Monash University},
             city={Clayton},
             postcode={3168}, 
            state={Victoria},
             country={Australia}}

\author[1]{Lixiang Yan}

\author[1]{Vanessa Echeverria}

\author[1]{Dragan Gašević}

\author[1]{Roberto Martinez-Maldonado}






\cortext[1]{Corresponding author}

\fntext[1]{Correspondence to Yueqiao Jin, Centre for Learning Analytics at Monash, Faculty of Information Technology, Monash University, Clayton, Victoria, 3168, Australia. Email: ariel.jin@monash.edu}



\begin{abstract}
Integrating generative AI (GAI) into higher education is crucial for preparing a future generation of GAI-literate students. Yet a thorough understanding of the global institutional adoption policy remains absent, with most of the prior studies focused on the Global North and the promises and challenges of GAI, lacking a theoretical lens. This study utilizes the Diffusion of Innovations Theory to examine GAI adoption strategies in higher education across 40 universities from six global regions. It explores the characteristics of GAI innovation, including compatibility, trialability, and observability, and analyses the communication channels and roles and responsibilities outlined in university policies and guidelines. The findings reveal a proactive approach by universities towards GAI integration, emphasizing academic integrity, teaching and learning enhancement, and equity. Despite a cautious yet optimistic stance, a comprehensive policy framework is needed to evaluate the impacts of GAI integration and establish effective communication strategies that foster broader stakeholder engagement. The study highlights the importance of clear roles and responsibilities among faculty, students, and administrators for successful GAI integration, supporting a collaborative model for navigating the complexities of GAI in education. This study contributes insights for policymakers in crafting detailed strategies for its integration.
\end{abstract}




\maketitle

\section{Introduction}

The adoption of generative AI (GAI) in higher education has the potential to transform various educational practices in learning, teaching, and assessment \citep{yan2023practical,kasneci2023chatgpt}. Emerging research has identified GAI's diverse capabilities, including providing comprehensive feedback \citep{dai2023can}, exceeding the performance on reflective writing of the average student \citep{li2023can}, enhancing multimedia learning \citep{vartiainen2023using}, and pioneering the development of adaptive educational content \citep{mazzoli2023enhancing}. Despite these advancements, concerns regarding the digital divide have surfaced, highlighting how unequal access to such technologies might deepen educational disparities \citep{pontual2020applications}. Moreover, the potential reliance on GAI raises questions about its influence on students' critical thinking, creativity, and independence \citep{darvishi2023impact,yan2024genai}. In light of these challenges and opportunities, the role of institutional policies becomes critical in navigating the integration of GAI within higher education \citep{tsai2017learning,tsai2018sheila,cheng2024examining}.

Following the public release of ChatGPT, many universities initially adopted a cautious, wait-and-see approach \citep{wang2023seeing,moorhouse2023generative,cheng2024examining}. However, as GAI tools (e.g., ChatGPT and Midjourney) have become increasingly accessible to students, the necessity for well-defined guidelines and policies has become apparent. These policies are crucial for guiding the integration of GAI into curriculum development, assessment design, and upholding academic integrity \citep{xiao2023waiting,plata2023emerging}. While recent research has begun to explore these policies in relation to how institutions strategise the adoption of GAI \citep{moorhouse2023generative,wang2023seeing,plata2023emerging,sullivan2023chatgpt}, such analyses often lack a theoretical grounding. This tendency can lead to a narrow focus on the benefits and challenges of GAI, as depicted in institutional policies, while neglecting other critical dimensions. Notably, aspects like innovation characteristics, communication channels, and the assignment of roles and responsibilities within the adoption process have been underexplored attention in prior studies \citep{mcdonald2024generative,sullivan2023chatgpt,xiao2023waiting}. These key components are essential for a nuanced understanding of how new technologies are adopted in higher education \citep{tsai2017learning,tsai2018sheila}. Additionally, prior studies have predominantly focused on universities in the \textit{Global North}  (a term often used to describe wealthier, industrialised countries), resulting in findings that may not be representative of a broader, global perspective \citep{wang2023seeing,moorhouse2023generative}. 

This study aims to bridge the gaps identified above by analysing the adoption policies of GAI within higher education across 40 universities from six global regions, grounded in the Diffusion of Innovations Theory (DIT) (see Section \ref{sec:diff-innov-theory} for details)  \citep{rogers2014diffusion}.  DIT provides a robust framework for understanding how new technologies, such as GAI, are adopted and integrated within complex systems like higher education institutions. It offers insights into the factors influencing the adoption process, enabling a systematic analysis of how innovations spread across different contexts and cultures. Specifically, this study investigates the characteristics of GAI innovation, including compatibility, trialability, and observability. The analysis also examines the use of communication channels and delineates the roles and responsibilities outlined in university policies and guidelines. This approach not only enhances our theoretical understanding of GAI adoption but also provides a comprehensive view of the dynamics influencing GAI integration in diverse educational settings globally.

\section{Background and Related Work}

\subsection{Generative AI Policies in Higher Education} 

Recent studies have explored university policies, guidelines, and media coverage to understand the adoption and response to GAI. Top universities in the U.S. and globally have been analysed for their approach to integrating generative AI in education, showing a cautious yet open strategy with concerns about ethics, accuracy, and privacy \citep{mcdonald2024generative,moorhouse2023generative,wang2023seeing}. 
\citet{plata2023emerging} examined academic integrity articles and policies at leading global universities, suggesting a model for maintaining academic integrity in the GAI era. \citet{sullivan2023chatgpt} investigated the impact of ChatGPT on higher education in Australia, New Zealand, the U.S., and the U.K. through news articles. A case study by \citet{cheng2024examining} documented policy adaptations and management strategies at eight Hong Kong public universities using local newspapers. \citet{xiao2023waiting} conducted a quantitative analysis of the top 500 universities worldwide, focusing on their ChatGPT policies, revealing differences and factors influencing these policies. However, these studies mainly focus on the Global North or specific regions, lacking a comprehensive global perspective. Additionally, most research was completed by mid-2023, and with the rapid evolution of GAI and policies \citep{cheng2024examining}, continuous monitoring and re-evaluation are essential. Notably, there is a gap in theoretical grounding in these investigations.

\subsection{Diffusion of Innovations Theory} 
\label{sec:diff-innov-theory}

The Diffusion of Innovations Theory (DIT) offers a framework to understand how new ideas, practices, or objects, known as innovations, spread across different cultures and social systems \citep{rogers2014diffusion}. Essential in sociology, communication, and public health, DIT outlines the mechanisms and factors influencing the diffusion rate and breadth of innovations \citep{wejnert2002integrating}. It examines the complex interaction among four key elements: innovation, communication channels, the social system, and time. This theory is particularly useful for analysing the adoption and integration of GAI technologies in higher education, providing insights into the factors that affect their spread across educational institutions worldwide \citep{frei2022using,pinho2021application}. For instance, DIT can elucidate the process through which GAI-enhanced educational technologies become accepted, highlighting the importance of compatibility with educational objectives and effective dissemination of benefits across the institution. This understanding can inform the creation of policy guidelines designed to maximise the benefits of this new technology while minimising its potential risks. The current study focuses on innovation, communication channels, and the social system, due to the challenges in capturing temporal dynamics, especially the difficulty in accessing archived policy documents, as new policies often replace older ones, hindering temporal analysis.

\subsubsection{Innovation Characteristics.} Innovation is a core concept in DIT, encompassing novel ideological, practical, and technological advancements \citep{rogers2014diffusion}. GAI represents a significant technological innovation that could also drive ideological and practical changes in higher education \citep{chiu2023impact,swiecki2022assessment,yan2023practical}. DIT suggests that an innovation's adoption likelihood is influenced by five characteristics: relative advantage, compatibility, complexity, trialability, and observability \citep{rogers2014diffusion}. \textit{Relative advantage} looks at the potential improvements GAI offers in education. \textit{Compatibility} assesses how well GAI fits within the existing educational framework, with institutions more likely to adopt GAI if it aligns with their goals and curricula. \textit{Complexity} relates to the ease of adoption and the need for specific enhancements. \textit{Trialability} is the extent to which institutions can experiment with GAI before committing to its full implementation, reducing risks. \textit{Observability} focuses on the visibility of the innovation's benefits, where clear success metrics and shared results can encourage wider adoption by demonstrating GAI's value. Previous studies have already offered thorough discussions on the concepts of relative advantage and complexity \citep{plata2023emerging,wang2023seeing}, with UNESCO's recent in-depth review in its guidance for GAI in education and research \citep{miao2023guidance} also contributing significantly to this area. However, the aspects of compatibility, trialability, and observability in the context of GAI policies within higher education remain under-researched. Understanding these three aspects of GAI's innovation characteristics could provide valuable insights into how higher education institutions are approaching its integration and the strategies they are developing or implementing for experimenting with and assessing GAI technologies' potential impacts on their educational systems. This identified gap has led to the formulation of the first research question:

\begin{itemize}
    \item \textbf{RQ1:} \textit{How are GAI's innovation characteristics, specifically compatibility, trialability, and observability, represented in higher education institutions' policies and guidelines?}
\end{itemize}

\subsubsection{Communication Channels.} Communication Channels play a crucial role in the diffusion of innovations, as highlighted by DIT \citep{rogers2014diffusion}. These channels, ranging from mass media to interpersonal networks, significantly influence perceptions and attitudes towards new innovations \citep{moolenaar2014linked}. In higher education, the significance of communication channels becomes even more pronounced, requiring a two-way flow of information. This approach facilitates the active participation of all key stakeholders, including faculty, students, and administrators, in the decision-making process, thus increasing the chances of successful innovation adoption \citep{gasevic2019we,tsai2017learning,tsai2018sheila}. The integration of GAI into higher education highlights the need for clear communication channels, particularly in light of concerns about academic integrity, assessment reform, and data privacy \citep{rudolph2023chatgpt,winograd2023loose}. However, the presence and effectiveness of these channels within the policy frameworks of educational institutions are not well-documented. This gap led to the second research question:

\begin{itemize}
    \item \textbf{RQ2:} \textit{What communication channels are identified in higher education policies for disseminating updates and facilitating discussions on GAI adoption among stakeholders?}
\end{itemize}

\subsubsection{Social System.} The Social System refers to the norms, structures, and the community's readiness to adopt new innovations within a given diffusion context \citep{rogers2014diffusion}. In higher education, this system is characterised by the institutional culture, policy environment, the network of educators and researchers, and the broader educational framework \citep{pinho2021application}. The policies set by educational institutions play a crucial role in demonstrating the system's readiness to adopt new technologies. They do this by defining the roles and responsibilities of key stakeholders in the process of adopting GAI \citep{kamal2011analyzing,okai2020readiness}. Despite its importance, there has been limited research on how higher education policies address the adoption of GAI from the perspective of the social system. This research gap is the focus of the last research question:

\begin{itemize}
    \item \textbf{RQ3:} \textit{What roles and responsibilities are specified for faculty, students, and administrators in higher education policies regarding the adoption of GAI?}
\end{itemize}

\section{Methods}
\subsection{Data Collection} 

To provide a comprehensive global perspective and avoid a focus solely on the Global North, this study utilised the QS World University Rankings 2024\footnote{\url{https://www.topuniversities.com/world-university-rankings}}, which categorises universities into six regions: Africa, Asia, Europe, Latin America, North America, and Oceania. Using stratified sampling \citep{neyman1992two}, the top 10 universities from each region were selected, resulting in a total of 60 universities. Publicly available information was manually collected from the official websites of these universities using initial search terms such as "Generative AI or ChatGPT" and "policy or guideline or guidance or guide." Additional terms identified during the search, such as "artificial intelligence or AI tools" and "learning or teaching or assessment," were included in subsequent searches to improve data collection comprehensiveness and rigour.

To overcome language barriers, searches were conducted in English and the official languages of the universities. Bilingual researchers performed searches in Spanish and Chinese for universities in Latin America and China (Mainland), respectively. In other regions, searches were conducted in English and the universities' primary languages, with translation assistance. All documents found were translated into English for analysis. The selection criteria for the search results were: 1) The document must be an official statement, policy, guideline, or guidance, excluding blogs, commentaries, news items, training courses, and memos. 2) The document must be issued at the university level, ensuring institution-wide applicability. 3) The document must primarily support teaching, learning, and assessment activities, excluding those intended for other purposes like academic research and information security.

The search concluded on 20th January 2024. After applying the selection criteria, 41 out of the 60 universities (68\%) had published relevant documents on their websites. The National University of Singapore was excluded due to website accessibility issues. Among the remaining 40 universities, 10 were in Oceania, nine in North America, eight in Europe, six in Africa, four in Asia, and three in Latin America. The dataset, including the list of universities, and archived web pages and documents, is available at the provided \href{https://osf.io/bj95p/?view_only=a9c63524fd0d43108aa795e69afa27b2}{\textit{link}}.

\subsection{Analysis} 

To address the research questions (RQ1--3), we conducted a thematic analysis to identify relevant themes from the documents \citep{braun2019reflecting}. This method allowed for an in-depth examination of the innovation characteristics, communication channels, and social systems related to GAI adoption policies at universities. Our goal was to offer insights that could assist policymakers in developing comprehensive GAI adoption strategies. In accordance with best practices for conducting thematic analysis \citep{braun2021one}, our initial step was to allow themes to naturally emerge from our dataset. To achieve this, two researchers independently reviewed the same set of 40 policy documents, identifying emerging themes pertinent to each RQ. Subsequently, these researchers convened to discuss their findings, merging similar themes to create a comprehensive codebook for each RQ (refer to Table \ref{table-rq1}, \ref{table-rq2}, and \ref{table-rq3} for further details). Utilising this codebook, the researchers proceeded to independently code the ten policy documents. Inter-rater reliability was measured using Cohen's Kappa for each identified theme. Themes with a Cohen's Kappa score below 0.61, indicative of less than substantial agreement \citep{mchugh2012interrater}, were subjected to further discussion. This entailed refining the descriptions of such themes and re-coding them by both researchers until a substantial agreement (Cohen's Kappa > 0.61) was achieved. The inter-rater reliability scores for each theme have been documented in the tables.

For \textbf{RQ1--Innovation Characteristic}, the analysis focused on compatibility, trialability, and observability as defined in DIT \citep{rogers2014diffusion}. We identified themes related to \textit{compatibility}, reflecting how universities perceive the integration of GAI as aligning with their goals and principles. For \textit{trialability}, themes were identified that highlight the approaches universities intend to use for experimenting with and incrementally implementing GAI tools in education. Regarding \textit{observability}, we looked for themes describing the methods and procedures established for evaluating the impact and effectiveness of GAI integration within the university environment. For \textbf{RQ2--Communication Channels}, themes were extracted about the various communication channels mentioned in the policy documents for disseminating updates and fostering dialogue on GAI adoption among stakeholders. Concerning \textbf{RQ3--Role and Responsibility}, themes were separately identified for faculty, students, and administrators based on their defined roles and responsibilities.

{
\fontsize{7pt}{9pt}\selectfont
\renewcommand{\arraystretch}{2}
\begin{longtable}{
  p{\dimexpr.1\textwidth-2\tabcolsep} 
  p{\dimexpr.2\textwidth-2\tabcolsep} 
  p{\dimexpr.275\textwidth-2\tabcolsep} 
  p{\dimexpr.275\textwidth-2\tabcolsep} 
  p{\dimexpr.05\textwidth-2\tabcolsep} 
  p{\dimexpr.1\textwidth-2\tabcolsep} 
}
\caption{Innovation Characteristics} \\
\toprule
\textbf{Characteristics} & \textbf{Themes} & \textbf{Description} & \textbf{Example} & \textbf{N} & \textbf{Kappa} \\
\midrule
\endfirsthead

\multicolumn{6}{c}%
{\tablename\ \thetable\ -- \textit{Continued from previous page}} \\
\toprule
\textbf{Characteristics} & \textbf{Themes} & \textbf{Description} & \textbf{Example} & \textbf{N} & \textbf{Kappa} \\
\midrule
\endhead

\midrule
\multicolumn{6}{r}{\textit{Continued on next page}} \\
\endfoot

\bottomrule
\endlastfoot

Compatibility & Academic Integrity and Ethical Use of AI & Universities highlight the importance of using Generative AI tools in a manner that maintains academic integrity and aligns with the principles of originality and honesty. & "Content produced by AI does not represent original content generated by a student and will be considered a form of academic misconduct if used indiscriminately or inappropriately." [University of Cambridge] & 40 & 1.00 \\
 & Enhancing Teaching and Learning & Universities see GAI and ChatGPT as tools to increase the effectiveness of teaching methodologies, enhance learning experiences, improve educational outcomes, and foster continuous learning. & "Instead of seeing this as a threat, let it serve as a wake-up call for higher education leaders and faculty to take a fresh look at pedagogical approaches and assessment tools." [Nanyang Technological University] & 38 & 0.79 \\
 & Fostering AI Literacy and Skills for the Future & Universities emphasize the development of AI literacy and the ability to use AI tools responsibly in professional settings, aiming to prepare students for a future where AI plays a significant role. & "At Oxford we want to support students and teaching staff to use AI ethically and appropriately in their work, and to regard AI literacy as a key skill... Universities will support students and staff to become AI-literate."[University of Oxford] & 33 & 0.68 \\
 & Information Security and Data Privacy & This theme focuses on information security, data privacy, compliance with existing policies, and the safe and secure use of AI technologies. & "At present, any use of ChatGPT should be with the assumption that no personal, confidential, proprietary, or otherwise sensitive information may be used with it." [University of California, Berkeley] & 25 & 0.83 \\
 & Support for Diverse Educational Needs and Equity & This theme encompasses personalized learning, support for students with disabilities, and ensuring equitable access to educational tools. & "Another affordance concerns the capacity of AI to absorb indigenous languages datasets. At present, the availability of AI in African languages is scarce at present, ... The more languages are shared on AI platforms, the more sophisticated the capacity becomes, in any language, and so what is anticipated is a fair(er) balance between participation in AI generation between the Global North and the Global South." [North-West University] & 10 & 0.82 \\
Trialability & Integrating AI into Educational Practice & This theme emphasizes the importance of making AI a part of the educational curriculum, encouraging its use for enhancing content creation, facilitating interactive learning, and adapting teaching and assessment strategies to leverage AI for educational benefits. & "In this regard, it is suggested to indicate the value that AI can add to certain tasks...Cleaning and querying for problems in programming codes. AI can provide quick responses if we ask about errors that may occur when programming a function...Brainstorming on topics for research projects or others related to the course."(Translated) [Pontificia Universidad Católica de Chile] & 40 & 1.00 \\
 & Critical Evaluation and Human-Centric Perspectives and Competencies & Universities encourages a critical stance towards AI-generated content and prioritize human-centric perspectives and skills, including the verification of information for accuracy and encouraging students to critically evaluate AI tool outputs. & "Then the students' assignment is to engage with ChatGPT's response, such as by fact-checking it against multiple academic sources, critiquing its responses, evaluating its responses from their vantage point in the Global South or in a particular context, or identifying possible biases, with the ChatGPT response included as part of students'submissions." [the University of Witwatersrand] & 39 & 0.66 \\
 & Transparency and Privacy & This theme emphasizes the need for transparency and privacy protection in using AI tools in education, covering the disclosure of AI involvement in content creation and assessment, guidelines for citing AI-generated content, respecting privacy, avoiding personal or confidential data use, and discussing ethical implications with students. & "Report and cite any use you make of ChatGPT or any other artificial intelligence system to perform academic work. In no case do you present, in whole or in part, the work of someone else or an artificial intelligence tool as if it were your own." [Tecnológico de Monterrey]; "...staff must not submit student work into such tools as this may compromise students' intellectual property and personal data rights." [University College London] & 38 & 0.64 \\
 & Policy Communication & This theme addresses the need for clear policy communication regarding AI use, ensuring that faculty and students are aware of guidelines and expectations. & "Whether you’re allowing the use of AI tools for certain assignments or banning them altogether, you and your students will benefit from a clear statement of what role AI should play in the class." [University of Chicago] & 36 & 0.88 \\
 & Assessment Design and Authentic Assessments & This theme emphasizes designing assessments that evaluate higher-order thinking skills and are resistant to being solved solely by AI, thereby ensuring academic rigor. & "...it may now be necessary to target forms of knowledge and expression that are more difficult for generative AI technologies - critical thinking, evaluation or creativity" and "Rather than only assessing the final output or 'product' of students' work, it is possible to also assess the process through which they produce it." [Monash University] & 35 & 0.72 \\
Observability & Continuous Evaluation & Universities are planning to test and evaluate the effectiveness of AI tools through new assessment methods and periodic evaluations involving various stakeholders. & "HKU will conduct periodic evaluations to assess whether its AI tools are achieving their intended outcomes in terms of enhancing T\&L (teaching and learning) ... The results of these evaluations will be made publicly available to ensure transparency." [University of Hong Kong] & 5 & 0.80 \\
 & Collaboration and Discussion & Universities are focusing on fostering collaboration and dialogue among educators, students, and other stakeholders to address the challenges and opportunities presented by AI. & "This challenge will be addressed through dialogue among educators in every field, transcending academic boundaries ... We plan to provide forums for university-wide discussions on this issue, and we hope to exchange ideas and situations with people from every academic discipline." [University of Tokyo] & 4 & 0.68 \\
 & Ongoing Monitoring & Universities are committing to continuously monitor AI technologies, adapting their use cases, policies, and guidance as their understanding and the technologies themselves evolve. & "This page will remain a work-in-progress and will be updated as use cases and engagement with ChatGPT technology continues to evolve." [University of California, Berkeley] & 3 & 0.72 \\
\label{table-rq1}
\end{longtable}
}

{
\fontsize{8pt}{10pt}\selectfont
\renewcommand{\arraystretch}{2}
\begin{longtable}{
  p{\dimexpr.25\textwidth-2\tabcolsep} 
  p{\dimexpr.6\textwidth-2\tabcolsep} 
  p{\dimexpr.05\textwidth-2\tabcolsep} 
  p{\dimexpr.1\textwidth-2\tabcolsep} 
}
\caption{Communication Channels} \\
\toprule
\textbf{Communication Channels} & \textbf{Description} & \textbf{N} & \textbf{Kappa} \\
\midrule
\endfirsthead

\multicolumn{4}{c}%
{\tablename\ \thetable\ -- \textit{Continued from previous page}} \\
\toprule
\textbf{Communication Channels} & \textbf{Description} & \textbf{N} & \textbf{Kappa} \\
\midrule
\endhead

\midrule
\multicolumn{4}{r}{\textit{Continued on next page}} \\
\endfoot

\bottomrule
\endlastfoot

Digital Platforms & This category includes official university websites, dedicated pages for AI guidance for faculty and students, public webpages, and blogs. & 15 & 0.80 \\
Interactive Learning and Engagement Channels & This category includes live, interactive sessions such as webinars and workshops, as well as forums and discussion panels coordinated by university offices or departments. & 9 & 0.81 \\
Direct and Personalized Communication Channels & This category covers direct communication channels, email communications, and video conferencing. It includes email updates, direct contact with privacy offices or IT services, memos to teaching faculty, and facilitating discussions via video conferencing platforms. & 7 & 0.78 \\
Collaborative and Social Networks & This category refers to the development of new social and professional networks, chat groups and dedicated channels within collaboration platforms aimed at facilitating conversations about AI among faculty, students, and other stakeholders. & 3 & 1.00 \\
Advisory, Monitoring, and Feedback Channels & This category consists of committees or teams formed to guide decisions on the incorporation of AI tools into teaching and learning environments such as advisory and working groups with monitoring and feedback mechanisms. & 3 & 0.84 \\
\label{table-rq2}
\end{longtable}
}

{
\fontsize{8pt}{10pt}\selectfont
\renewcommand{\arraystretch}{2}
\begin{longtable}{
  p{\dimexpr.1\textwidth-2\tabcolsep} 
  p{\dimexpr.2\textwidth-2\tabcolsep} 
  p{\dimexpr.55\textwidth-2\tabcolsep} 
  p{\dimexpr.05\textwidth-2\tabcolsep} 
  p{\dimexpr.1\textwidth-2\tabcolsep} 
}
\caption{Roles and Responsibilities} \\
\toprule
\textbf{Roles} & \textbf{Responsibility} & \textbf{Description} & \textbf{N} & \textbf{Kappa} \\
\midrule
\endfirsthead

\multicolumn{5}{c}%
{\tablename\ \thetable\ -- \textit{Continued from previous page}} \\
\toprule
\textbf{Roles} & \textbf{Responsibility} & \textbf{Description} & \textbf{N} & \textbf{Kappa} \\
\midrule
\endhead

\midrule
\multicolumn{5}{r}{\textit{Continued on next page}} \\
\endfoot

\bottomrule
\endlastfoot

Faculty & Integrating GAI into Curriculum and Assessment & This theme involves faculties incorporating Generative Artificial Intelligence (GAI) technologies into the curriculum and assessment methods. It reflects efforts to adapt teaching and learning methodologies to include GAI tools, thereby enhancing educational frameworks. & 20 & 0.90 \\
 & Communication and Education on GAI Use & This theme emphasizes the importance of faculties in communicating clearly about GAI applications and educating students on the careful and responsible use of GAI and detection tools. & 17 & 0.84 \\
 & Setting Guidelines and Policies & Faculties are tasked with establishing ethical standards, operational protocols, and clear guidelines for GAI usage within their courses. This involves creating a framework within which GAI tools can be used responsibly and effectively. & 9 & 0.78 \\
 & Enhancing Critical Thinking & Faculties should explore the use of GAI to improve pedagogical outcomes by fostering environments that cultivate intellectual growth and enhance students' critical thinking abilities. & 4 & 0.88 \\
 & Reviewing Ethical and Security Concerns & Faculties are responsible for navigating potential risks associated with GAI integration, including concerns related to ethics, privacy, and security. & 3 & 0.64 \\
Student & Ethical Use and Academic Integrity & Students are expected to use GAI tools ethically and responsibly, and maintain academic integrity. This theme includes appropriately acknowledging the use of GAI tools in their academic work, complying with university policies, and ensuring responsible use of data. & 27 & 0.94 \\
 & Understanding the capabilities and limitations of GAI & Students are encouraged to understand the strengths and weaknesses of GAI technology and develop a critical perspective towards it and content generated by these tools. & 6 & 0.75 \\
 & Engagement and Communication & This theme highlights the importance for students to actively participate in discussions about the ethical and practical implications of GAI in education. & 6 & 0.77 \\
 & Learning Enhancement & Students are expected to actively adopt GAI tools and leverage its power in supporting and enhancing the learning experience. & 4 & 0.91 \\
Administrator & Policy Development and Implementation & This theme involves activities related to creating, updating, and implementing policies and guidelines to govern the use and ensure the effective integration of GAI tools within the university. It includes establishing codes of conduct, advisory committees, and best practices. & 16 & 0.95 \\
 & Support and Resources & This theme focuses on offering support and resources to faculty and students to guide the proper use of GAI tools and foster AI literacy. It includes providing resources, technical support, and fostering dialogue around GAI. & 16 & 0.80 \\
 & Academic Integrity and Ethical Use & This theme encompasses measures to ensure academic integrity and the ethical use of GAI tools. It includes detecting improper use of GAI tools and managing breaches of academic integrity. & 7 & 0.78 \\
 & Supervision in Procurement Process of GAI Services & This theme is specifically focused on the supervision of purchasing GAI tools to minimize institutional risk. It includes overseeing the acquisition of such tools and ensuring they align with university policies and ethical standards. & 3 & 0.84 \\
\label{table-rq3}
\end{longtable}
}

\section{Results}

\subsection{RQ1--Compatibility} 
\begin{figure}[htbp]
    \centering
      \includegraphics[width=.95\textwidth]{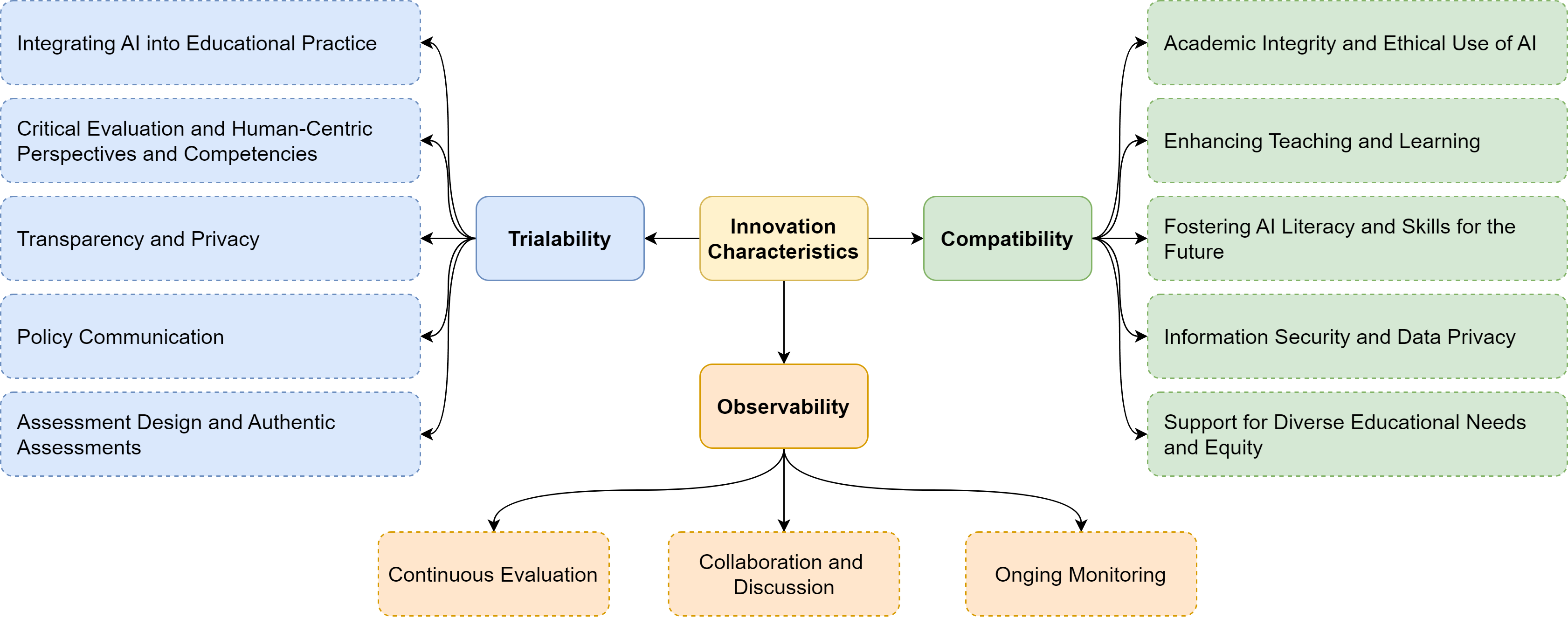}
    \caption{Key themes related the compatibility, trialability, observability of generative AI integration which emerged from the analysed universities' policies and guidelines.}
    \label{fig-rq1}
\end{figure}

In the adoption of GAI across 40 universities, as shown in Figure \ref{fig-rq1}, five key themes emerged to align the technology with their existing institutional goals and objectives. The most common theme highlighted by all the universities is \textbf{academic integrity and ethical use of AI} (n=40). These universities suggested the potential incompatibility between GAI tools and the principles of \textit{originality} and \textit{honesty}. For instance, the University of Cambridge (United Kingdom) and the University of Sydney (Australia) warned against the inappropriate use of AI-generated content, considering it a form of academic misconduct. Similarly, Stanford University (United States) treated the use of generative AI analogously to assistance from another person without a clear statement from the instructor. Moreover, the University of Cape Town (South Africa) highlighted the concerns around plagiarism and cheating from inappropriate use of GAI tools in its guides for both teaching staff and students.

The alignment between GAI and universities' objectives to \textbf{enhance teaching and learning} (n=38) was identified. Universities see GAI as a tool to increase the effectiveness of teaching methodologies, enhance learning experiences, improve educational outcomes, and foster continuous learning. Specifically, Nanyang Technological University (Singapore) viewed GAI not as a threat but as an opportunity to re-evaluate pedagogical approaches and assessment tools, aiming to adapt its educational practices to ongoing advancements in GAI. Likewise, the Chinese University of Hong Kong (China Hong Kong SAR) encouraged teachers and students to embrace and utilise appropriate AI tools to enhance their teaching and learning experiences. The University of Johannesburg (South Africa) also emphasised the role of GAI as a great learning assistant to provide personalised support across various contexts, including practical sessions, tutorials, and collaborative learning. 

The adoption of GAI is also being viewed as compatible with the objective of universities to cultivate essential skills that are vital for preparing students for the future workforce. Specifically, universities aimed to prepare students for a future where AI plays a significant role, with an focus on \textbf{fostering AI literacy and skills} for responsible use in professional settings (n=33). For example, the University of Oxford (United Kingdom), emphasised the goal on developing AI literacy as an essential skill for students, highlighting its commitment to preparing them for a future where AI is ubiquitous. The University of Hong Kong (China Hong Kong SAR) has recognised the importance of GAI literacy alongside traditional academic literacies and committed an action plan for developing GAI literacy among both students and instructors. A self-paced online module will be made available to students, with future plans to make it a credit-bearing course. For instructors, the university encourages the exploration of GAI in pedagogy, provides targeted online training, and offers personalised support through its GAI Task Force and Teaching and Learning Innovation Center. Likewise, the Universidad Nacional Autónoma de México (Mexico) stressed educating students about AI to foster a critical attitude towards its use and encourage critical evaluation of these tools' outputs, further demonstrating the universities' commitment to fostering AI literacy as an essential skill for students.

A potential incompatibility was noted between GAI and some universities' requirements on \textbf{information security and data privacy} (n=25). For instance, the University of California, Berkeley (United States), emphasised the importance of assuming that no personal, confidential, proprietary, or otherwise sensitive information may be used with ChatGPT. Likewise, the University of Cape Town highlighted the ethical and legal risks associated with sharing student data by uploading it into detectors managed by third-party companies with unknown privacy and data usage policies. The Universidad Nacional Autónoma de México also indicated the challenges and responsibilities universities face in ensuring data privacy and security when using AI-based applications. For instance, the university is responsible for providing users with detailed information about data usage, and ensuring that data collection is necessary and conforms to reasonable expectation.

Lastly, universities noted the double-edged sword nature of GAI tools in achieving the goal of \textbf{supporting diverse educational needs and promoting equity} (n=10) and were taking active measures to mitigate this potential incompatibility. In one example, Cornell University (United States) encouraged using AI along with inclusive instructional approaches like Universal Design for Learning (UDL) to enhance course accessibility. When integrating a GAI tool in teaching, the University of California, Berkeley, advocated for alternative assignment options if the specified GAI tool is inaccessible to students, particularly for those with disabilities. Similarly, North-West University (South Africa) highlighted the importance of language and cultural inclusion by leveraging AI to support indigenous languages. Financial considerations were prioritised by the California Institute of Technology (United States), the University of Manchester (United Kingdom), and the University of Hong Kong, focusing on providing equitable access to GAI tools without additional costs to students, ensuring that no student is disadvantaged due to financial constraints. 

\subsection{RQ1--Trialability}

Regarding universities' strategic exploration into the experimental and phased integration of GAI tools within educational frameworks, we identified five key themes (Figure \ref{fig-rq1}). The most prevalent theme was to encourage the \textbf{integration of AI into educational practices} (n=40), including content creation, interactive learning, and teaching and assessment strategies. For example, Pontificia Universidad Católica de Chile (Chile) encouraged students' use of GAI for quick responses in programming code issues and brainstorming research topics, showcasing the practical applications of GAI in enhancing course-specific skills. From a broader pedagogical perspective, the University of Cape Town provided nine use cases for both staff and students to explore the potential benefits for deployment in teaching and learning, such as content creation, learning analytics, personalised learning support, and automated grading systems. In contrast, the University of Queensland (Australia) imposed stricter rules, requiring educators to choose a specific level of GAI use permitted in their courses and to clearly declare this in the course profile.

Universities' integration strategy and guidelines also focused on \textbf{critical evaluation and human-centric perspectives and competencies} (n=39). For example, Cornell University offered guidance on leveraging GAI platforms to help students become more discerning and engage with GAI as a tutor to enhance critical thinking. Likewise, the University of Witwatersrand (South Africa) proposed a new form of assignment where students actively engage with AI-generated responses by verifying them against academic sources, and assessing them from their unique perspectives in the Global South. North-West University highlighted its mission to raise awareness of the potential limitation of GAI, noting that human agents remain indispensable for fostering development and growth in education.

Phased implementation and experimentation of GAI also co-occurred with regulatory exploration into \textbf{transparency and privacy} (n=38), ensuring the openness of GAI tools in educational contexts. For instance, Tecnológico de Monterrey (Mexico) advised students to verify the correct use of AI technology with their professors and to cite any use of GAI systems in academic work. Similarly, ETH Zurich (Switzerland) suggested citing the prompt used for generating the content to maintain academic integrity which would aid the establishment of norms for the citation of AI-generated content. On the other hand, University College London (United Kingdom) emphasised the importance of not uploading personal data to AI systems without considering data protection requirements, highlighting the need for approvals and consents overseen by a staff supervisor. Likewise, the University of Edinburgh (United Kingdom) advised treating information given to AI tools as if it were posted on a public site. This advice aimed to bring caution against sharing personal, confidential, copyrighted, or restricted information, preventing privacy issues and the potential for misuse of AI tools. 

The importance of \textbf{clear policy communication} regarding GAI (n=36) during phased implementation and GAI experimentation was also identified. The University of Auckland (New Zealand) considers communication with students to be "critical" for managing their expectations. Therefore, instructors need to be explicit about the permitted tools to prevent students from unintentionally engaging in academic misconduct. Similarly, the University of Chicago (United States) advocated for a clear statement in class syllabi about the role of GAI, offering guidance for syllabus statements on GAI tool usage to benefit both instructors and students. The University of Sydney published guidelines on the use of GAI in education, research, and operations, while recommended open discussions with students about GAI. This approach was echoed by the University of Edinburgh's plans to establish a site for sharing information on GAI use and a group to update and share best practices.

Lastly, universities also encouraged instructors to conduct trials and experiments with novel \textbf{assessment designs and authentic assessments} (n=35) that target high-order cognitive abilities and can remain impervious to GAI solutions. For instance, the University of Adelaide (Australia) emphasised the importance of designing assessments that require genuine student effort and creativity, rather than tasks that can be easily completed by GAI. Similarly, Monash University (Australia) and the University of Pretoria (South Africa) advised instructors on modifying questions to be more applicable to individual student contexts and incorporating real-life problem-solving into assessments, examining students' applied knowledge and their critical thinking and analytical skills. Monash University also emphasised the necessity of assessing the process rather than merely the final output of students' work.

\subsection{RQ1--Observability} The results reveal that seven universities were adopting a proactive approach towards the integration and monitoring of GAI technologies within their ecosystems, focusing on three main themes: continuous evaluation, collaboration and discussion, and ongoing engagement and updates (Figure \ref{fig-rq1}). Firstly, \textbf{continuous evaluation} (n=5) is a part of universities' strategies to monitor the outcomes and effectiveness of GAI integration. The University of New South Wales (Australia) and the University of Hong Kong, for example, were actively testing new assessment methods and conducting periodic evaluations with multiple stakeholders to assess the impact of AI tools on teaching and learning. These evaluations aimed to ensure transparency and adaptability in the use of AI within educational settings.

Secondly, \textbf{collaboration and discussion} (n=4) emerged as an evaluation approach across institutions such as the University of Tokyo (Japan) and Cornell University. The University of Tokyo highlighted the necessity of dialogue among instructors across various fields to address the challenges presented by AI, planning to provide forums for university-wide discussions. Similarly, Cornell University encouraged collaborative review of course assessment plans with colleagues and the Center for Teaching Innovation, emphasising the importance of faculty consultations on teaching and learning in the context of AI. Lastly, some universities (n=3) emphasised the importance of \textbf{ongoing monitoring} with AI technologies, promising continuous adaptation to their policies and guidance as their understanding of these technologies evolves. For instance, the University of California, Berkeley committed to regularly adapting the work-in-progress guidance as new use cases emerge and further engagement with the GAI technology develops. 

These themes reflect a comprehensive strategy for integrating GAI in university ecosystems, balancing innovation with critical assessment and collaboration to navigate the challenges and opportunities presented by these technologies.

\subsection{RQ2--Communication Channels} 
\begin{figure}[htbp]
    \centering
      \includegraphics[width=.75\textwidth]{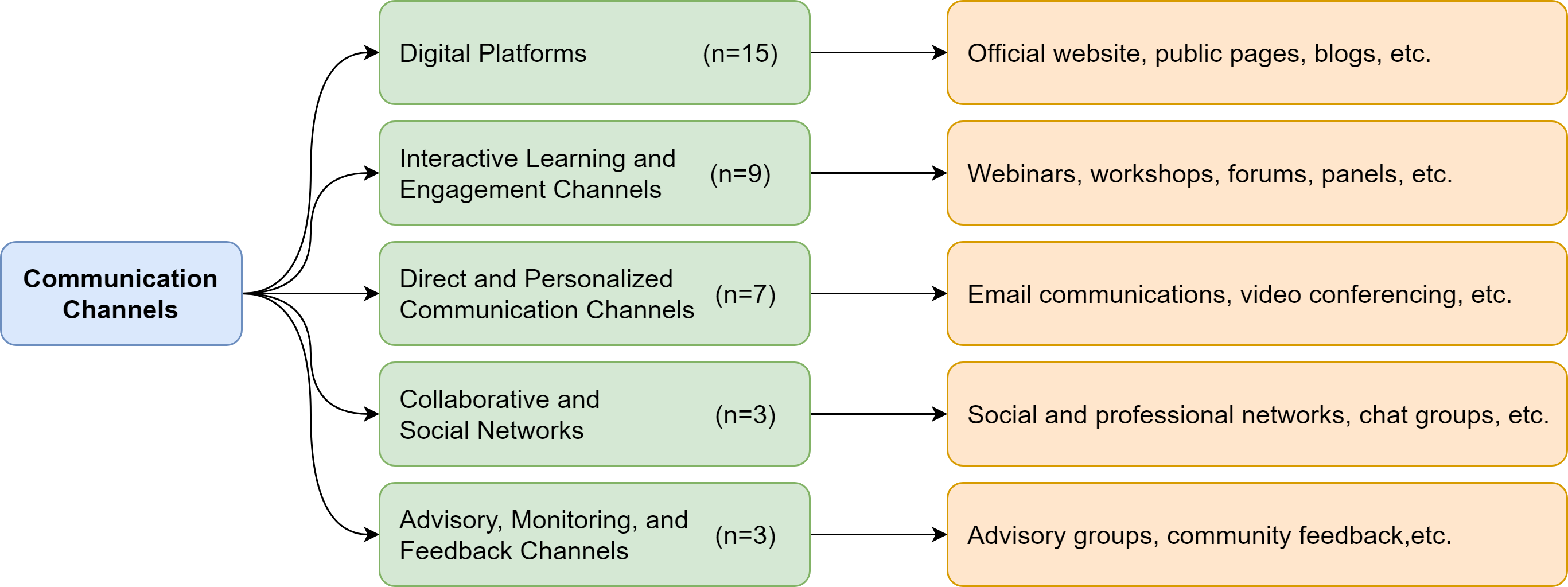}
    \caption{Five primary communication channels utilised by higher education institutions to communication information about the adoption policy of generative AI.}
    \label{fig-rq2}
\end{figure}

In addressing RQ2, our analysis identified a diverse range of communication channels within higher education policies (22/40 universities) aimed at conveying updates and stimulating conversations about GAI adoption among stakeholders. As shown in Figure \ref{fig-rq2}, these channels are broadly categorised into five main groups. Firstly, \textbf{digital platforms} (n=15) emerged as the most prevalent channel, comprising official university websites, AI guidance pages, and blogs. For example, the University of Cape Town utilised its official website for policy and guidance document updates, serving as a central repository for accessible information dissemination. Similarly, the University of Edinburgh planned to host an information-sharing site, providing a platform for the latest research and best practices in GAI. 

Secondly, \textbf{interactive learning and engagement channels} (n=9), including webinars, workshops, forums, and discussion panels, were designed to foster live, interactive sessions. The University of Witwatersrand, for instance, coordinated webinars and workshops to engage stakeholders in discussions on AI in teaching and learning, while the University of Sydney hosted staff forums and student panels to explore the impact of GAI on education. Thirdly, \textbf{direct and personalised communication channels} (n=7), such as email communications, direct contacts, and video conferencing, offered stakeholders a direct line for inquiries. The University of California, Berkeley, provides direct contact for privacy concerns and hosts workshops, whereas Princeton University (United States) facilitated Zoom discussions for teaching faculty, allowing for personalised advice and engagement on AI tools. 

Additionally, \textbf{collaborative and social networks} (n=3), including social and professional networks, collaboration platforms, and chat groups, supported real-time communication. The University of New South Wales leveraged Microsoft Teams groups for faculty members to share information about respective news, academic literature, workshops, and events. Lastly, \textbf{advisory, monitoring, and feedback channels} (n=3), through advisory committees and feedback processes, played a crucial role in guiding AI tool incorporation decisions. The University of Hong Kong, for example, planned to establish an advisory committee to decide which GAI tools should be incorporated into their teaching and learning environment, ensuring ethical use and collecting community feedback to inform policy and practice. 

These findings highlight the multifaceted approach higher education institutions are adopting to communicate about GAI, emphasizing the importance of diverse and interactive channels in fostering an informed and engaged academic community.

\subsection{RQ3--Role and Responsibility} 
\subsubsection{Faculties}

\begin{figure}[htbp]
    \centering
      \includegraphics[width=.75\textwidth]{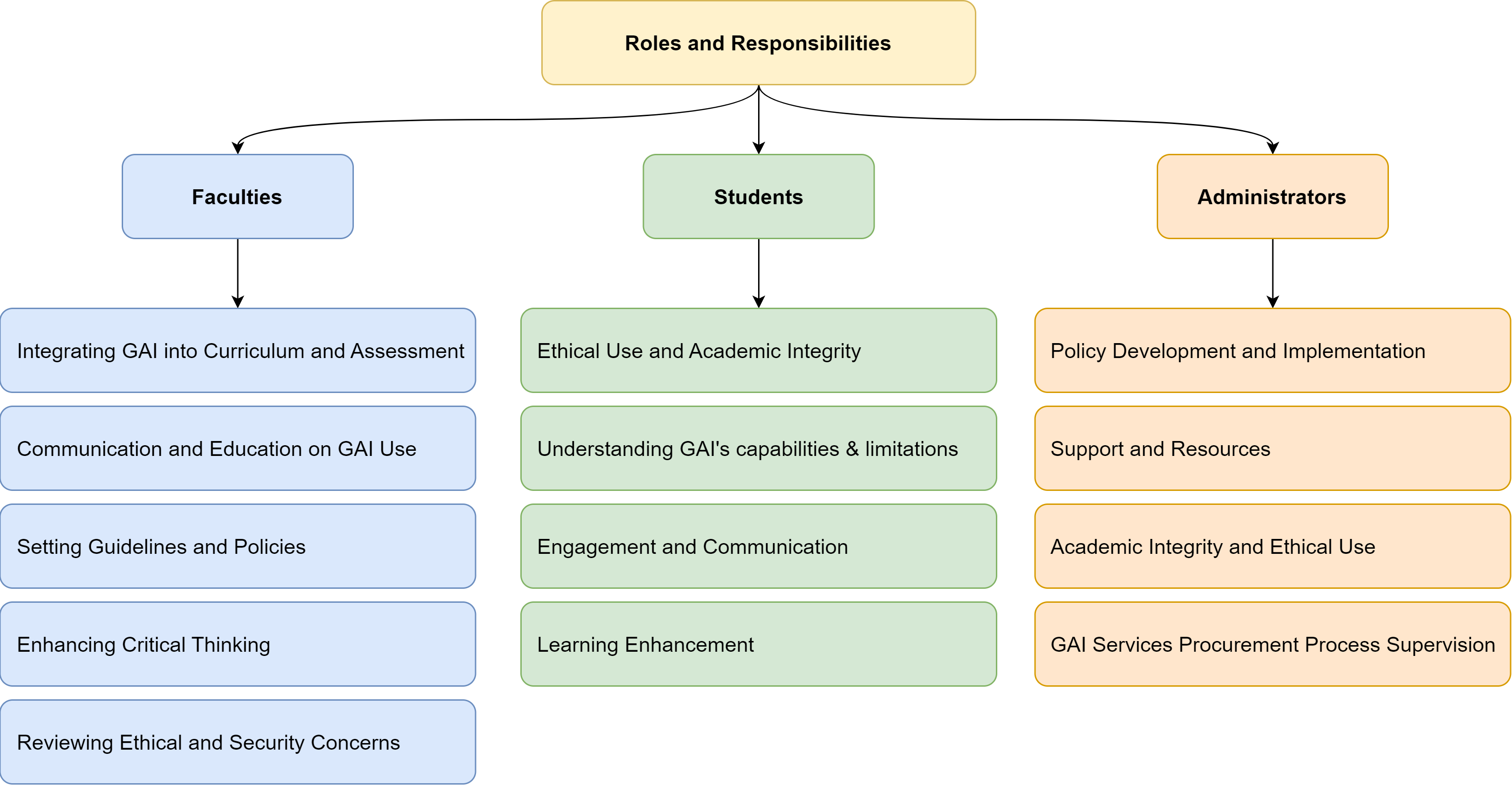}
    \caption{Roles and Responsibilities of faculties, students, and administrators in the adoption process of generative AI.}
    \label{fig-rq3}
\end{figure}

As shown in Figure \ref{fig-rq3}, we identified five distinct roles and responsibilities assigned to faculty across 36 institutions. Faculties were advised to \textbf{integrate GAI into curriculum and assessment} (n=20), reflecting efforts to incorporate GAI technologies within educational frameworks, thereby transforming teaching and learning methodologies. Specifically, the University of Melbourne encouraged faculties leveraging GAI tools to improve various aspects of assessment design, which indicates a strategic incorporation of GAI into curriculum and evaluation. Another major faculty role involved \textbf{communicating and educating students on GAI use} (n=17), highlighting the importance of clear communication about GAI applications and the careful use of GAI detection tools. The University of Cape Town, for example, accentuated the responsibility for instructors to clarify the appropriate use of GAI tools at the beginning of the courses. Furthermore, \textbf{setting guidelines and policies} was identified as a critical responsibility (n=9) by some universities, such as the University of Pennsylvania (United States) and ETH Zurich, with faculties being advised to focus on establishing ethical standards and operational protocols that aligns with the Principles of Responsible Conduct for GAI usage for their courses. The role of \textbf{enhancing students' critical thinking} (n=4) emphasised the strategic use of GAI to improve pedagogical outcomes and foster environments that cultivate intellectual growth. For example, The University of Pretoria provided recommendations for faculty to utilise ChatGPT to enhance students’ critical thinking abilities, suggesting the use of GAI as a tool to foster deeper intellectual engagement and problem-solving skills. Lastly, attention to \textbf{reviewing ethical and security concerns} (n=3) highlighted the role of navigating potential risks, including accessibility, data privacy, and security, associated with GAI integration.

\subsubsection{Students}

Students were primarily charged with the duty of ensuring \textbf{ethical and responsible use of GAI tools and maintaining academic integrity} (n=27). The documents from these universities frequently mentioned several best practices for honesty and transparency within academic settings. These include acknowledging the use of AI tools, complying with university policies and course instructions, and exercising caution in data usage when deploying these tools. Additionally, the responsibilities extended to \textbf{understanding the capabilities and limitations of GAI} (n=6), which is crucial for fostering a critical perspective towards technology among students. The University of Edinburgh detailed this into three aspects, in which students are expected to understand the limitations of AI systems, check the factual accuracy of AI-generated content, and avoid over-reliance on AI tools as a single key source. The roles of \textbf{learning enhancement} (n=6) and \textbf{engagement and communication} (n=4) reflected a proactive stance towards the adoption of GAI in educational contexts, encouraging students to leverage GAI to support their learning experience and participate in meaningful discussions regarding the ethical and practical implications of GAI in education. Specifically, the Technical University of Munich (Germany) elaborated on the scenarios where students can use ChatGPT to support their learning processes or to use the chatbot as a peer, proofreader, or brainstorming partner for learning enhancement. In addition, the University of Sydney has actively engaged students in collaborating with staff to develop a dedicated Canvas site that serves as a hub for AI-related educational resources.

\subsubsection{Administrators}

Administrators were primarily tasked with \textbf{policy development and implementation} (n=16), a responsibility that encompasses a range of activities, including the establishment of conduct codes, advisory committees, and best practices. These activities aimed to facilitate the creation, updating, and implementation of policies, as well as the effective integration of GAI tools within university processes. A significant focus was also placed on \textbf{providing guidance and support} to both faculty and students (n=16), aiming to foster AI literacy and guide the proper use of GAI tools. For instance, the Centre for Learning and Teaching at the American University in Cairo (Egypt) has been instrumental in integrating AI tools. It serves as a central repository and contact point for faculty and students. Moreover, administrators play a crucial role in ensuring \textbf{academic integrity and ethical use of GAI} (n=7), through measures such as detecting improper use of GAI tools and managing breaches of integrity. Notably, the motivation to detect misuse of these tools is distinct from deploying AI detection tools themselves. For instance, Imperial College London has committed to staff training to enhance the understanding and identification of AI usage, prompted by concerns about the maturity and accuracy of AI detection tools. Lastly, administrators were also tasked with the responsibility of \textbf{supervising GAI tool procurement} to minimise institutional risk (n=3). For instance, at Yale University, efforts to support procurement practices that align shared interests and minimise institutional risks exemplify the supervision of GAI tool procurement. This ensures that the adoption of these technologies adheres to the university's ethical standards and risk management policies.

\section{Discussion}

In our study, we examined GAI adoption policies in higher education, analysing data from 40 universities across six global regions through the Diffusion of Innovations Theory (DIT) framework. For research question one (\textbf{RQ1}), we focused on the innovation characteristics of GAI, particularly its compatibility, trialability, and observability within higher education policies. Our findings highlighted a universal concern for academic integrity and the ethical use of AI, underscoring institutions' dedication to maintaining core educational values amidst technological progress \citep{moorhouse2023generative,wang2023seeing}. The positive perception of GAI as a tool to enhance teaching and learning effectiveness indicated a strategic shift towards leveraging technology to meet educational objectives, highlighting its potential to complement and augment traditional teaching methodologies \citep{xiao2023waiting,cheng2024examining}. Moreover, the \textbf{compatibility} between GAI and the goal of fostering innovation and 21st-century skills stressed the need to integrate these tools into educational practices, ensuring that institutions remain relevant in an AI-influenced future \citep{yan2024genai,darvishi2023impact}. Although less common, a focus on promoting equity revealed a growing effort among some institutions to mitigate inequalities and ensure GAI benefits a broad student demographic, preventing further educational disparities \citep{pontual2020applications}. 

Our analysis revealed a strategic focus on \textbf{trialability}, with higher education institutions actively promoting the integration of GAI into educational practices. This included emphasising critical evaluation, human-centric competencies, and advocating for a phased implementation that explores regulatory aspects of transparency and privacy. These findings highlighted the proactive approach of universities in leveraging GAI's potential while addressing its ethical and practical challenges. Universities were actively integrating GAI into their curricula and pedagogical strategies, recognising its transformative potential to enhance educational outcomes, foster critical thinking, and prepare students for future careers. The focus on critical evaluation and human-centric competencies suggested a balanced view of GAI as a complement to human intellect and creativity, rather than a replacement \citep{yan2024genai,darvishi2023impact}. This approach aligns with findings from previous studies that analysed documents, such as policies and new articles, from both quantitative and qualitative perspectives \citep{xiao2023waiting,cheng2024examining}.

The \textbf{observability} of GAI integration within higher education, as revealed by our findings, underscores a nuanced landscape where a subset of universities is taking significant strides toward embedding GAI technologies within their educational ecosystems. The identified themes of ongoing engagement, collaboration, and continuous evaluation reflect a proactive and iterative approach to adopting GAI, highlighting the importance of adaptability, transparency, and community involvement. However, the relatively small number of universities (n=7) actively engaging in evaluation measures for GAI's impact signals a gap in the comprehensive policy development and implementation across the higher education sector. This gap not only raises questions about the readiness of the majority of institutions to adapt to and fully leverage GAI technologies but also points to potential disparities in the resources available for such endeavors, such as institutional infrastructures and staff skills \citep{king2015exploring}. The limited engagement suggests a need for a more widespread and structured approach to understanding and harnessing the benefits and challenges of GAI, as well as a more robust framework for sharing best practices and learning across institutions \citep{cheng2024examining,kezar2011best}. Moreover, the emphasis on continuous evaluation, collaboration, and ongoing monitoring by a minority of universities (n=5, n=4, and n=3, respectively) raises critical questions about the scalability and sustainability of such practices. While these intended approaches are commendable, their impact and effectiveness may be limited if they are not adopted more broadly across the higher education sector. This situation highlights the need for a more inclusive dialogue on GAI integration, one that encompasses a wider array of institutions, including those that may lack the resources or infrastructure to engage in such proactive measures \citep{soomro2020digital}. 

Regarding communication channels (\textbf{RQ2}), our analysis showed that approximately a half of the universities employed a varied approach to foster dialogue and share information among educational stakeholders. This strategy included digital platforms, interactive learning environments, direct communication methods, collaborative networks, and advisory mechanisms, highlighting the complex and multifaceted process of integrating GAI into education. The use of digital platforms for policy updates, alongside interactive and direct communication channels, can be a balanced strategy for broad outreach and individual engagement. This method is consistent with prior research advocating for active stakeholder participation and diverse communication channels to facilitate successful innovation adoption in higher education \citep{tsai2017learning,tsai2018sheila}. However, the fact that less than half of the universities implemented these channels indicated a critical need for more focused efforts in disseminating and communicating GAI adoption policies, given the technology's rapid evolution \citep{yan2023practical,kasneci2023chatgpt}. Moreover, a limited emphasis on collaborative and social networks (n=3) contrasts with literature highlighting their role in promoting a community-oriented, innovative climate \citep{moolenaar2014linked}, suggesting an area for enhancement.

For the final research question (\textbf{RQ3}), our findings revealed that the majority of universities (n=36) have established clear policies defining the roles and responsibilities of faculty, students, and administrators in GAI adoption. This structured approach facilitated the integration of GAI into educational practices. Faculty were tasked with incorporating GAI into teaching and ensuring its ethical use by students. Students were responsible for using GAI ethically and participating in critical discussions about its implications. Administrators were charged with crafting and implementing policies that ensure GAI's integration aligns with institutional values and academic integrity. 

While previous studies have examined the roles of faculty and students in GAI adoption \citep{moorhouse2023generative,wang2023seeing}, our research provided a more detailed perspective by also considering administrators' roles, offering insights into the comprehensive stakeholder responsibilities in GAI's context. Our findings highlighted the importance of a collaborative effort among all educational ecosystem stakeholders to effectively manage GAI integration's complexities. This approach suggested that successful GAI adoption in higher education depends not just on the technology but also on the community's preparedness and involvement. This perspective supports existing literature on the significance of clearly defined roles and responsibilities in adopting educational technologies \citep{kamal2011analyzing,okai2020readiness}, emphasising the need for a structured approach to fully leverage GAI's potential.

The current study has several \textbf{limitations}, including potential sample selection bias towards well-resourced institutions due to reliance on QS World University Rankings, language and cultural interpretation challenges despite efforts to include bilingual researchers, and the dynamic nature of GAI policies development, which may have evolved beyond the study's cutoff date. Similarly, choosing the top 10 universities from each region may not accurately reflect the situation of all universities globally, particularly those with limited resources. Future research could benefit from qualitative investigations involving educational stakeholders at these less-resourced institutions to better comprehend their strategies for adopting GAI. Such investigations could be critical to understand the impacts of GAI adoption on the digital divide and existing inequalities within the higher education sector. Lastly, the focus on official policy documents excludes informal communications and may not fully capture the practical implementation of GAI or the perspectives of key stakeholders like students and faculty. Future research could benefit from a broader sample, qualitative insights from direct stakeholders, and ongoing updates to remain relevant.

\section{Conclusion}

This study identified the strategic and proactive measures higher education institutions globally are adopting towards the integration of GAI, guided by a commitment to academic integrity, teaching, learning enhancement, and equity. Our analysis, framed by the DIT, revealed a universal emphasis on GAI's compatibility with educational values, its potential for fostering innovation and critical thinking, and the importance of trialability and observability in its adoption. Despite the optimism surrounding GAI's capabilities, the findings also highlighted significant gaps in comprehensive policy development, communication strategies, and the equitable distribution of resources for GAI integration. The identified need for a structured approach to stakeholder engagement, especially in defining clear roles and responsibilities, underscores the necessity of a collaborative and inclusive model that navigates the complexities of GAI adoption in education, ensuring its alignment with institutional values and the broader educational mission.











\bibliographystyle{cas-model2-names}

\bibliography{0_reference}



\end{document}